\documentclass{elsart}

\usepackage{cite}

\usepackage{graphicx}

\begin{document}

\begin{frontmatter}

\title{Non-hermitean hamiltonians with unitary and antiunitary symmetry}
\author{Francisco M. Fern\'{a}ndez\thanksref{FMF}} \and \author{Javier Garcia}

\address{INIFTA (UNLP, CCT La Plata-CONICET), Divisi\'on Qu\'imica Te\'orica,
Blvd. 113 S/N,  Sucursal 4, Casilla de Correo 16, 1900 La Plata,
Argentina}

\thanks[FMF]{e--mail: fernande@quimica.unlp.edu.ar}

\begin{abstract}
We analyse several non-Hermitian Hamiltonians with antiunitary symmetry from
the point of view of their point-group symmetry. It enables one to predict
the degeneracy of the energy levels and to reduce the dimension of the
matrices necessary for the diagonalization of the Hamiltonian in a given
basis set. We can also classify the solutions according to the irreducible
representations of the point group. One of the main results of this paper is
that PT-symmetric Hamiltonians with point-group symmetry $C_{2v}$ exhibit
complex eigenvalues for all values of a potential parameter. In such cases
the PT phase transition takes place at the trivial Hermitian limit and
suggests that the phenomenon is not robust.
\end{abstract}


\end{frontmatter}

\section{Introduction}

\label{sec:intro}

It was shown some time ago that some complex non-Hermitian Hamiltonians may
exhibit real eigenvalues\cite{CGM80, A95}. The conjecture that such
intriguing feature may be due to unbroken PT-symmetry\cite{BB98} gave rise
to a very active field of research\cite{B07} (and references therein). The
first studied PT-symmetric models were mainly one-dimensional anharmonic
oscillators\cite{BB98,B07,FGZ98,FGRZ99} and lately the focus shifted towards
multidimensional problems\cite{BDMS01,NA02,N02,N05,BTZ06,W09,CIN10,BW12,HV13}%
. Among the most widely studied multidimensional PT-symmetric models we
mention the complex versions of the Barbanis\cite
{BDMS01,NA02,N05,BTZ06,W09,BW12,HV13} and H\'{e}non-Heiles\cite{BDMS01,W09}
Hamiltonians. Several methods have been applied to the calculation of their
spectra: the diagonalization method\cite{BDMS01,NA02,N02,N05,W09,BW12},
perturbation theory\cite{BDMS01,N02,N05,W09}, classical and semiclassical
approaches\cite{BDMS01,NA02}, among others\cite{W09,HV13}. Typically, those
models depend on a potential parameter $g$ so that the Hamiltonian is
Hermitian when $g=0$ and non-Hermitian when $g\neq 0$. Bender and Weir\cite
{BW12} conjectured that some of those models may exhibit PT phase
transitions so that their spectra are entirely real for sufficiently small
but nonzero values of $|g|$. Such phase transition appears to be a
high-energy phenomenon.

Multidimensional oscillators exhibit point-group symmetry (PGS)\cite
{PE81a,PE81b}. As far as we know such a property has not been taken into
consideration in those earlier studies of the PT-symmetric models, except
for the occasional parity in one of the variables. It is to be expected that
PGS may be relevant to the study of the spectra of multidimensional
PT-symmetric anharmonic oscillators. One of the purposes of this paper is to
start such research.

The main interest in the study of PT-symmetric oscillators has been to
enlarge the class of such models that exhibit real spectra, at least for
some values of the potential parameter. In such cases PT-symmetry is broken
at particular values $g=g_{c}$ of the parameter that are known as
exceptional points\cite{HS90,H00,HH01,H04} and can be easily calculated as
critical parameters by means of the diagonalization method\cite{FG13}. The
PT phase transition is determined by the smallest $|g_{c}|$. Another goal of
this paper is to find PT-symmetric models that do not exhibit real spectra,
except at the trivial Hermitian limit $g=0$.

In section~\ref{sec:AU_symmetry} we outline the main ideas of unitary
(point-group) and antiunitary symmetry. In section~\ref{sec:solvable} we
show that two exactly solvable PT-symmetric oscillators with different PGS
exhibit quite different spectra. In sections \ref{sec:Barbanis}, \ref
{sec:H-H} and \ref{sec:H3D} we discuss some non-Hermitian operators, already
studied earlier by other authors, from the point of view of PGS. All of them
have been shown to exhibit high-energy phase transitions. In section~\ref
{sec:present} we show a PT-symmetric anharmonic oscillator with complex
eigenvalues for all values of the potential parameter. Finally, in section~%
\ref{sec:conclusions} we summarize the main results of the paper and draw
conclusions.

\section{Unitary and antiunitary symmetry}

\label{sec:AU_symmetry}

We assume that there is a group of unitary transformations $%
G=\{U_{1},U_{2},\ldots ,U_{n}\}$ and a set of antiunitary transformations $%
S=\{A_{1},A_{2},\ldots ,A_{m}\}$ that leave the non-Hermitian Hamiltonian
operator invariant
\begin{equation}
U_{j}HU_{j}^{-1}=H,\;A_{j}HA_{j}^{-1}=H.  \label{eq:invariance}
\end{equation}
Therefore, if $\psi $ is an eigenvector of $H$ with eigenvalue $E$ we have
\begin{equation}
HU_{j}\psi =EU_{j}\psi ,  \label{eq:HUpsi}
\end{equation}
and
\begin{equation}
\;HA_{j}\psi =E^{*}A_{j}\psi .  \label{eq:HApsi}
\end{equation}
It follows from the latter equation that the eigenvalues of $H$ are either
real or appear as pairs of conjugate complex numbers.

It is well known that a product of antiunitary operators is a unitary one%
\cite{W60}. Therefore, since $A_{i}A_{j}$ leaves the Hamiltonian invariant
then $A_{i}A_{j}=U_{k}$ $\in G$, provided that $G$ is the actual symmetry
point group for $H$\cite{T64,C90}.

If $A_{j}\psi =\lambda \psi $ then the antiunitary symmetry is said to be
unbroken and $E=E^{*}$. For some non-Hermitian Hamiltonians with degenerate
states the eigenvalue can be real even though $A_{j}\psi \neq \lambda \psi $%
\cite{FG13}.

\section{Exactly solvable examples}

\label{sec:solvable}

In this section we discuss exactly solvable PT-symmetric models similar to
those studied earlier by Nanayakkara\cite{N02} and Cannata et al\cite{CIN10}%
. In the present case we focus on the PGS of the Hamiltonian operators that
was not considered by those authors. As a first simple model we consider the
Hamiltonian operator
\begin{equation}
H=p_{x}^{2}+p_{y}^{2}+x^{2}+y^{2}+iaxy  \label{eq:H_solv_1}
\end{equation}
where $a$ is a real parameter. It is exactly solvable and invariant under
the operations of the symmetry point group $C_{2v}$: $\{E,C_{2},\sigma
_{v1},\sigma _{v2}\}$ that transform the variables according to
\begin{eqnarray}
E &:&(x,y)\rightarrow (x,y),  \nonumber \\
C_{2} &:&(x,y)\rightarrow (-x,-y),  \nonumber \\
\sigma _{v1} &:&(x,y)\rightarrow (y,x),  \nonumber \\
\sigma _{v2} &:&(x,y)\rightarrow (-y,-x).
\end{eqnarray}
Note that $C_{2}$ is a rotation by an angle $\pi $ around the $z$ axis and $%
\sigma _{v}$ are vertical reflection planes\cite{T64,C90}. Unless otherwise
stated, from now on it is assumed that the same transformations apply to the
momenta $(p_{x},p_{y})$. In the case of a two-dimensional model the effect
of the symmetry operations on the $z$ variable is irrelevant and for this
reason there may be more than one point group suitable for the description
of the problem. For example, here we can also choose the symmetry point
groups $C_{2h}$or $D_{2}$\cite{T64,C90}. For concreteness we restrict
ourselves to the $C_{2v}$ point group with irreducible representations $%
\{A_{1},\,B_{1},\,A_{2},\,B_{2}\}$.

To the PGS discussed above we can also add the antiunitary operations
\begin{equation}
A(x)=C_{2}(x)T,\;A(y)=C_{2}(y)T,
\end{equation}
where $T$ is the time reversal operation\cite{P65} and
\begin{eqnarray}
C_{2}(x) &:&(x,y)\rightarrow (x,-y),  \nonumber \\
C_{2}(y) &:&(x,y)\rightarrow (-x,y),
\end{eqnarray}
are rotations by $\pi $ about the $x$ and $y$ axis, respectively. Note that $%
A(x)A(y)=C_{2}$ is an example of the product of two antiunitary operators
that results in one of the elements of the symmetry point group for $H$.

This model is separable into two harmonic oscillators by means of the change
of variables
\begin{eqnarray}
x &=&\frac{1}{\sqrt{2}}(s+t),  \nonumber \\
y &=&\frac{1}{\sqrt{2}}(s-t),
\end{eqnarray}
that leads to
\begin{eqnarray}
H &=&p_{s}^{2}+p_{t}^{2}+ks^{2}+k^{*}t^{2},  \nonumber \\
k &=&1+i\frac{a}{2}.
\end{eqnarray}
If we write $\omega =\sqrt{k}=\omega _{R}+i\omega _{I}$ then the eigenvalues
are given by
\begin{equation}
E_{mn}=2(m+n+1)\omega _{R}+2(m-n)i\omega _{I},
\end{equation}
where $m,n=0,1,\ldots $ and
\begin{equation}
\omega _{R}=\sqrt{\frac{1}{2}+\frac{1}{2}\sqrt{1+\frac{a^{2}}{4}}},\;\omega
_{I}=\frac{a}{4\omega _{R}}.
\end{equation}
We see that all the eigenvalues with $m=n$ are real and those with $m\neq n$
are complex when $a\neq 0$ (more precisely: $E_{mn}=E_{nm}^{*}$). In this
case the PT phase transition\cite{BW12} takes place at the trivial Hermitian
limit $a=0$. It is also obvious that the perturbation series for this model
exhibits only powers of $a^{2}$ when $m=n$ and powers of $a$ when $m\neq n$.

The eigenfunctions can be written as
\begin{equation}
\psi _{mn}(s,t)=\phi _{m}(k,s)\phi _{n}(k^{*},t),
\end{equation}
where $\phi _{m}(k,s)$ is an eigenfunction of $p_{s}^{2}+ks^{2}$. Therefore
\begin{eqnarray}
A(x)\psi _{mn}(s,t) &=&\psi _{mn}^{*}(t,s)=\psi _{nm}(s,t),  \nonumber \\
A(y)\psi _{mn}(s,t) &=&\psi _{mn}^{*}(-t,-s)=(-1)^{m+n}\psi _{nm}(s,t)
\end{eqnarray}
that are consistent with equation (\ref{eq:HApsi}).

The states $\psi _{2m\,2n}$, $\psi _{2m+1\,2n+1}$, $\psi _{2m+1\,2n}$ and $%
\psi _{2m\,2n+1}$ are bases for the irreducible representations $A_{1}$, $%
A_{2}$, $B_{1}$ and $B_{2}$, respectively. It is clear that only some of the
states with symmetry $A_{1}$ and $A_{2}$ have real eigenvalues and that
those with symmetry $B_{1}$ and $B_{2}$ exhibit only complex ones. Moreover,
the antiunitary operators $A(x)$ and $A(y)$ transform functions of symmetry $%
B_{1}$ into functions of symmetry $B_{2}$ and viceversa, which shows that PT
symmetry is broken for all $a\neq 0$. More precisely, the eigenvalue of $%
\psi _{2m+1\,2n}$ ($B_{1}$) is the complex conjugate of $\psi _{2n}\,_{2m+1}$
($B_{2}$).

We also appreciate that the eigenfunctions of the non-Hermitian Hamiltonian
retain their symmetry in the Hermitian limit: $\lim\limits_{a\rightarrow
0}\psi _{mn}(s,t)=\phi _{m}(1,s)\phi _{n}(1,t)$.

In order to test the effect of symmetry on the spectra of the non-Hermitian
Hamiltonians we next consider the less symmetric operator
\begin{equation}
H=p_{x}^{2}+p_{y}^{2}+2x^{2}+y^{2}+iaxy,
\end{equation}
that is invariant under the operations of the point group $C_{2}$: $%
\{E,C_{2}\}$. In this case all the eigenvalues
\begin{equation}
E_{mn}=(2m+1)\omega _{1}+(2n+1)\omega _{2},
\end{equation}
where
\begin{equation}
\omega _{1}=\sqrt{\frac{3}{2}+\frac{\sqrt{1-a^{2}}}{2}},\;\omega _{2}=\sqrt{%
\frac{3}{2}-\frac{\sqrt{1-a^{2}}}{2}},
\end{equation}
are real provided that $|a|<1$. In this less symmetric example we find a PT
phase transition at the exceptional point $a=1$. The eigenfunctions are
bases for the irreducible representations $\{A,B\}$.

From the results of this section we may argue that PGS determines whether
the PT symmetry is broken or unbroken. In order to confirm such conjecture
we should find other examples (preferably non exactly solvable) with PT
phase transitions at the trivial Hermitian limit. Before doing it we first
discuss the non-Hermitian Hamiltonians studied so far from the point of view
of PGS.

\section{The Barbanis Hamiltonian}

\label{sec:Barbanis}

The PT-symmetric version of the Barbanis Hamiltonian
\begin{equation}
H=\frac{1}{2}\left( p_{x}^{2}+p_{y}^{2}\right) +\frac{1}{2}\left(
x^{2}+y^{2}\right) +iaxy^{2},
\end{equation}
is one of the simplest nontrivial two-dimensional models chosen by several
authors as a suitable illustrative example\cite
{BDMS01,NA02,N05,BTZ06,W09,BW12,HV13}. Most of them have in fact exploited
the fact that it is invariant under $y$ parity: $P_{y}:(x,y)\rightarrow
(x,-y)$. If we take into account that the effect of $P_{y}$ is equivalent to
a rotation by an angle $\pi $ about the $x$ axis then we realize that the
appropriate symmetry point group for this model is $C_{2}$ with elements $%
\{E,C_{2}(x)\}$ and irreducible representations $\{A,B\}$\cite{T64,C90}.
This model with a rather low symmetry appears to exhibit a PT phase
transition at $a\approx 0.1$\cite{BW12}.

The slightly modified Hamiltonian\cite{BW12}
\begin{equation}
H=\frac{1}{2}\left( p_{x}^{2}+p_{y}^{2}\right) +\frac{1}{2}%
x^{2}+y^{2}+iax^{2}y,
\end{equation}
exhibits the same symmetry and in this case the phase transition occurs
approximately at $a\approx 0.08$.

\section{The H\'{e}non-Heiles Hamiltonian}

\label{sec:H-H}

A more interesting non-Hermitian anharmonic oscillator is the PT-symmetric
version of the H\'{e}non-Heiles one\cite{BDMS01,W09}
\begin{equation}
H=p_{x}^{2}+p_{y}^{2}+x^{2}+y^{2}+ia\left( xy^{2}-\frac{1}{3}x^{3}\right) .
\end{equation}
Earlier treatments of this problem have taken into account the $y$ parity
already discussed in the preceding section. This symmetry is insufficient to
account for the existence of two-fold degenerate eigenvalues already
mentioned by Wang\cite{W09}. The fact is that this Hamiltonian is invariant
under rotations around the $z$ axis by angles $2\pi /3$ and $4\pi /3$ as
well as under three vertical and equivalent reflection planes $\sigma _{v}$%
\cite{PE81b}. The appropriate symmetry point group is thus $C_{3v}$ and the
eigenfunctions are bases for the irreducible representations $%
\{A_{1},A_{2},E\}$\cite{T64,C90}. The latter is two-dimensional and accounts
for the degeneracy just mentioned.

If instead of the three vertical planes $\sigma _{v}$ we choose three
equivalent axes $C_{2}$ perpendicular to the principal $C_{3}$ one the
suitable point group results to be $D_{3}$. The results coming from any of
these choices are equivalent. In section~\ref{sec:solvable} we already
explained why we can choose more than one symmetry point group for the
two-dimensional models discussed here.

The eigenvalues and eigenfunctions of the Hermitian operator $H(a=0)$ are
\begin{equation}
E_{mn}(a=0)=2(m+n+1),\;m,n=0,1,\ldots ,
\end{equation}
and
\begin{equation}
\varphi _{mn}(x,y)=\phi _{m}(x)\phi _{n}(y),
\end{equation}
respectively, where $\phi _{j}(q)$ is a normalized eigenfunction of the
harmonic oscillator $H=p_{q}^{2}+q^{2}$. It is convenient for the discussion
below to label the eigenfunctions as $\psi _{M,j}(x,y)$, where $M=m+n$, $%
j=0,1,\ldots ,M$ and $E_{M,0}\leq E_{M,1}\leq \ldots \leq E_{M,M}$ so that
(as outlined in section~\ref{sec:solvable})
\begin{equation}
\lim\limits_{a\rightarrow 0}\psi
_{M,j}(x,y)=\sum_{i=0}^{M}c_{M-i,i,j}\varphi _{M-i,i}(x,y),
\end{equation}
where the coefficients $c_{ij}$ are determined by the symmetry of the
eigenfunction. For example, the first eigenfunctions in this limit and their
corresponding symmetries are
\begin{eqnarray}
M &=&0:\;\{\varphi _{00}\},\;A_{1},  \nonumber \\
M &=&1:\;\{\varphi _{10},\varphi _{01}\},\;E,  \nonumber \\
M &=&2:\;\left\{
\begin{array}{c}
\left\{ \frac{1}{\sqrt{2}}\left( \varphi _{20}+\varphi _{02}\right) \right\}
,\;A_{1}\; \\
\left\{ \frac{1}{\sqrt{2}}\left( \varphi _{20}-\varphi _{02}\right) ,\varphi
_{11}\right\} ,\;E
\end{array}
\right. .
\end{eqnarray}

The projection operators $P^S$ are suitable for a systematic construction of
symmetry-adapted functions\cite{T64,C90}. For example, for $M=3$ we have
\begin{eqnarray}
P^{A_{1}}\varphi _{30} &=&\frac{1}{4}\varphi _{30}-\frac{\sqrt{3}}{4}\varphi
_{12},  \nonumber \\
P^{A_{2}}\varphi _{21} &=&\frac{3}{4}\varphi _{21}-\frac{\sqrt{3}}{4}\varphi
_{03},  \nonumber \\
P^{E}\varphi _{30} &=&\frac{3}{4}\varphi _{30}-\frac{\sqrt{3}}{4}\varphi
_{12},  \nonumber \\
P^{E}\varphi _{21} &=&\frac{1}{4}\varphi _{21}+\frac{\sqrt{3}}{4}\varphi
_{03}.
\end{eqnarray}
These functions are not normalized to unity because $\left\langle P^S\varphi
\right| \left. P^S\varphi \right\rangle \leq \left\langle \varphi \right|
\left. \varphi \right\rangle $ for any projection operator $P^S$. Note that
the functions with symmetry $A_{1}$ and $A_{2}$ exhibit even and odd parity,
respectively, with respect to the operation $P_{y}$ discussed above. On the
other hand, one of the functions of the basis for the irreducible
representation $E$ is even and the other odd.

The order of the energy levels for this model when $a$ is sufficiently small
(say, $a=0.1$) is
\begin{equation}
\begin{array}{lll}
2(M+1) & \rm{Symmetry} & \rm´{Ref.}$\cite{W09}$ \\
2 & A_{1} & E_{00} \\
4 & E & E_{10}=E_{11} \\
6 & \left\{
\begin{array}{c}
E \\
A_{1}
\end{array}
\right. & \left\{
\begin{array}{l}
E_{21}=E_{22} \\
E_{20}
\end{array}
\right. \\
8 & \left\{
\begin{array}{c}
A_{1} \\
A_{2} \\
E
\end{array}
\right. & \left\{
\begin{array}{l}
E_{32} \\
E_{33} \\
E_{41}=E_{42}
\end{array}
\right. \\
10 & \left\{
\begin{array}{c}
E \\
E \\
A_{1}
\end{array}
\right. & \left\{
\begin{array}{l}
E_{43}=E_{44} \\
E_{41}=E_{42} \\
E_{40}
\end{array}
\right. \\
12 & \left\{
\begin{array}{c}
E \\
A_{1} \\
A_{2} \\
E
\end{array}
\right. & \left\{
\begin{array}{l}
E_{54}=E_{55} \\
E_{52} \\
E_{53} \\
E_{50}=E_{51}
\end{array}
\right.
\end{array}
\end{equation}
where the last column shows the energy levels as labelled by Wang\cite{W09}
who derived the perturbation expansions

\begin{eqnarray}
E_{00} &=&2+\frac{a^{2}}{18}-\frac{11a^{4}}{864}+\frac{6089a^{6}}{933120}-%
\frac{2221951a^{8}}{447897600}+\ldots  \nonumber \\
E_{10} &=&E_{11}=4+\frac{7}{18}a^{2}-\frac{133}{864}a^{4}+\frac{30191}{233280%
}a^{6}-\frac{67779467}{447897600}a^{8}+\ldots  \nonumber \\
E_{20} &=&6+\frac{31}{18}a^{2}-\frac{145}{288}a^{4}+\frac{200923}{186624}%
a^{6}-\frac{40752209}{29859840}a^{8}+\ldots  \nonumber \\
E_{21} &=&E_{22}=6+\frac{5}{9}a^{2}-\frac{83}{144}a^{4}+\frac{432493}{466560}%
a^{6}-\frac{133188257}{74649600}a^{8}+\ldots  \nonumber \\
E_{30} &=&E_{31}=8+\frac{26}{9}a^{2}-\frac{535}{432}a^{4}+\frac{180037}{46656%
}a^{6}-\frac{296084959}{44789760}a^{8}+\ldots  \nonumber \\
E_{32} &=&8+\frac{5}{9}a^{2}-\frac{1123}{432}a^{4}+\frac{1416869}{233280}%
a^{6}-\frac{3963323843}{223948800}a^{8}+\ldots  \nonumber \\
E_{33} &=&8+\frac{5}{9}a^{2}-\frac{115}{432}a^{4}+\frac{12121}{46656}a^{6}-%
\frac{15676999}{44789760}a^{8}+\ldots  \nonumber \\
E_{40} &=&10+\frac{91}{18}a^{2}-\frac{2065}{864}a^{4}+\frac{1208431}{186624}%
a^{6}-\frac{1731827209}{89579520}a^{8}+\ldots  \nonumber \\
E_{41} &=&E_{42}=10+\frac{35}{9}a^{2}-\frac{1085}{432}a^{4}+\frac{1285823}{%
93312}a^{6}-\frac{1478364167}{44789760}a^{8}+\ldots  \nonumber \\
E_{43} &=&E_{44}=10+\frac{7}{18}a^{2}-\frac{2485}{864}a^{4}+\frac{1063615}{%
186624}a^{6}-\frac{1819581169}{89579520}a^{8}+\ldots  \nonumber \\
E_{50} &=&E_{51}=12+\frac{127}{18}a^{2}-\frac{1205}{288}a^{4}+\frac{814129}{%
46656}a^{6}-\frac{1958220799}{29859840}a^{8}+\ldots  \nonumber \\
E_{52} &=&12+\frac{85}{18}a^{2}-\frac{2633}{288}a^{4}+\frac{1370563}{29160}%
a^{6}-\frac{20818356203}{149299200}a^{8}+\ldots  \nonumber \\
E_{53} &=&12+\frac{85}{18}a^{2}+\frac{55}{288}a^{4}+\frac{70673}{5832}a^{6}+%
\frac{354058961}{29859840}a^{8}+\ldots  \nonumber \\
E_{54} &=&E_{55}=12+\frac{1}{18}a^{2}-\frac{1457}{288}a^{4}+\frac{329257}{%
29160}a^{6}-\frac{9599275547}{149299200}a^{8}+\ldots
\end{eqnarray}

\section{Hamiltonian operator in three dimensions}

\label{sec:H3D}

Bender et al\cite{BDMS01} and Bender and Weir\cite{BW12} also
discussed some PT-symmetric Hamiltonians in three dimensions. One
of them is
\begin{equation}
H=p_{x}^{2}+p_{y}^{2}+p_{z}^{2}+x^{2}+y^{2}+z^{2}+iaxyz.
\label{eq:H3D}
\end{equation}
The eigenfunctions of $H_{0}=H(a=0)$ are given by
\begin{equation}
\varphi _{m,n,k}(x,y,z)=\phi _{m}(x)\phi _{n}(y)\phi
_{k}(z),\;m,n,k=0,1,\ldots ,
\end{equation}
and the corresponding eigenvalues $E_{mnk}^{(0)}=2M+3$,$\;M=m+n+k$, are $%
(M+1)(M+2)/2$-fold degenerate. The perturbation $H^{\prime
}=iaxyz$ splits the states of the three-dimensional Harmonic
oscillator $H_{0}$ in the following way:
\begin{eqnarray}
\{2n,2n,2n\} &\rightarrow &A_{1}  \nonumber \\
\{2n+1,2m,2m\}_{P} &\rightarrow &T_{2}  \nonumber \\
\{2n+1,2n+1,2m\} &\rightarrow &T_{2}  \nonumber \\
\{2n,2m,2m\} &\rightarrow &A_{1},E  \nonumber \\
\{2n+1,2n+1,2n+1\} &\rightarrow &A_{1}  \nonumber \\
\{2n,2m,2k+1\}_{P} &\rightarrow &T_{1},T_{2}  \nonumber \\
\{2n,2m+1,2k+1\}_{P} &\rightarrow &T_{1},T_{2}  \nonumber \\
\{2n,2m,2k\}_{P} &\rightarrow &A_{1},A_{2},E,E  \nonumber \\
\{2n+1,2m+1,2k+1\}_{P} &\rightarrow &A_{1},A_{2},E,E
\end{eqnarray}
where $\{m,n,k\}_{P}$ stands for all the distinct permutations of
the three positive integers. In this case the eigenvalues of
(\ref{eq:H3D}) appear to be real for all $0\leq
a<a_{c}$\cite{BW12}.
\section{Non-Hermitian oscillator with $C_{2v}$ point-group symmetry}

\label{sec:present}

In section~\ref{sec:solvable} we saw that the phase transition for the
exactly solvable example with symmetry point group $C_{2v}$ occurs at $a=0$.
The purpose of this section is to verify whether another PT-symmetric
anharmonic oscillator with that symmetry exhibits the same behaviour. A
suitable example is the non-Hermitian modification of the Pullen-Edmonds
Hamiltonian\cite{PE81a}
\begin{equation}
H=p_{x}^{2}+p_{y}^{2}+\alpha \left( x^{2}+y^{2}\right) +\beta
x^{2}y^{2}+iaxy.  \label{eq:H_present}
\end{equation}
Note that both the unitary and antiunitary transformations that leave this
Hamiltonian invariant are exactly those already introduced in section \ref
{sec:solvable}. In fact, when $\alpha =1$ and $\beta =0$ we obtain the first
exactly solvable example discussed there. When $a=0$ we recover the
Pullen-Edmonds Hamiltonian with $C_{4v}$ PGS\cite{PE81a}.

In order to discuss the results from the point of view of PGS we apply the
diagonalization method with symmetry-adapted products $\varphi _{mn}(x,y)$
of eigenfunctions $\phi _{n}(q)$ of the harmonic oscillator $%
H=p_{q}^{2}+q^{2}$. We thus obtain basis sets with the following functions
\begin{eqnarray}
\varphi _{2m\,2n}^{+} &=&\left\{
\begin{array}{c}
\varphi _{2n\,2n}(x,y),\;m=n \\
\frac{1}{\sqrt{2}}\left[ \varphi _{2m\,2n}(x,y)+\varphi
_{2n\,2m}(x,y)\right] ,\;m\neq n
\end{array}
\right. ,  \nonumber \\
\varphi _{2m\,2n}^{-} &=&\frac{1}{\sqrt{2}}\left[ \varphi
_{2m\,2n}(x,y)-\varphi _{2n\,2m}(x,y)\right] ,\;m\neq n,  \nonumber \\
\varphi _{2m+1\,2n+1}^{+} &=&\left\{
\begin{array}{c}
\varphi _{2n+1\,2n+1}(x,y),\;m=n \\
\frac{1}{\sqrt{2}}\left[ \varphi _{2m+1\,2n+1}(x,y)+\varphi
_{2n+1\,2m+1}(x,y)\right] ,\;m\neq n
\end{array}
\right. ,  \nonumber \\
\varphi _{2m+1\,2n+1}^{-} &=&\frac{1}{\sqrt{2}}\left[ \varphi
_{2m+1\,2n+1}(x,y)-\varphi _{2n+1\,2m+1}(x,y)\right] ,\;m\neq n,  \nonumber
\\
\varphi _{2m\,2n+1}^{+} &=&\frac{1}{\sqrt{2}}\left[ \varphi
_{2m\,2n+1}(x,y)+\varphi _{2n+1\,2m}(x,y)\right] ,  \nonumber \\
\varphi _{2m\,2n+1}^{-} &=&\frac{1}{\sqrt{2}}\left[ \varphi
_{2m\,2n+1}(x,y)-\varphi _{2n+1\,2m}(x,y)\right] ,
\end{eqnarray}
with symmetry
\begin{eqnarray}
\varphi _{2m\,2n}^{+},\;\varphi _{2m+1\,2n+1}^{+} &:&\;A_{1},  \nonumber \\
\varphi _{2m\,2n}^{-},\;\varphi _{2m+1\,2n+1}^{-} &:&\;A_{2},  \nonumber \\
\varphi _{2m\,2n+1}^{+} &:&\;B_{1},  \nonumber \\
\varphi _{2m\,2n+1}^{-} &:&\;B_{2}.
\end{eqnarray}
Since basis functions of different symmetry do not mix then we can carry out
four independent diagonalizations, one for each irreducible representation.

Because $A(x)\varphi _{2m\,2n+1}^{+}=-\varphi _{2m\,2n+1}^{-}$ then $%
A(x)\psi _{B_{1}}=\lambda _{B_{1}B_{2}}\psi _{B_{2}}$ and $%
E_{B_{1}}=E_{B_{2}}^{*}$ according to equation (\ref{eq:HApsi}). Therefore,
the eigenvalues for $B$ eigenfunctions are expected to be complex for any $%
a>0$ as in the case of the exactly solvable model discussed in section \ref
{sec:solvable}. We have verified this conclusion by numerical calculation
(see below).

Straightforward application of the diagonalization method with those
symmetry-adapted basis sets shows that there are no real eigenvalues with
eigenfunctions of symmetry $B$. More precisely, the characteristic
polynomials for the bases with symmetry $B_{1}$ and $B_{2}$ exhibit odd
powers of $g=ia$ which do not appear in those for the other two irreducible
representations $A_{1}$ and $A_{2}$. The characteristic polynomials for the
entire basis set $\{\varphi _{mn}\}$ are only functions of $g^{2}$ and the
complex eigenvalues appear as pairs of complex conjugate numbers. In other
words, the coefficients of the characteristic polynomials are real for the
full basis set as argued elsewhere\cite{F13}. On the other hand, the
coefficients of the characteristic polynomials for $B_{1}$ and $B_{2}$ are
complex and every complex root $E_{B_{1}}$ of the former has its counterpart
$E_{B_{2}}^{*}$ as a root of the latter.

\begin{figure}[tbp]
\begin{center}
\bigskip\bigskip\bigskip \includegraphics[width=6cm]{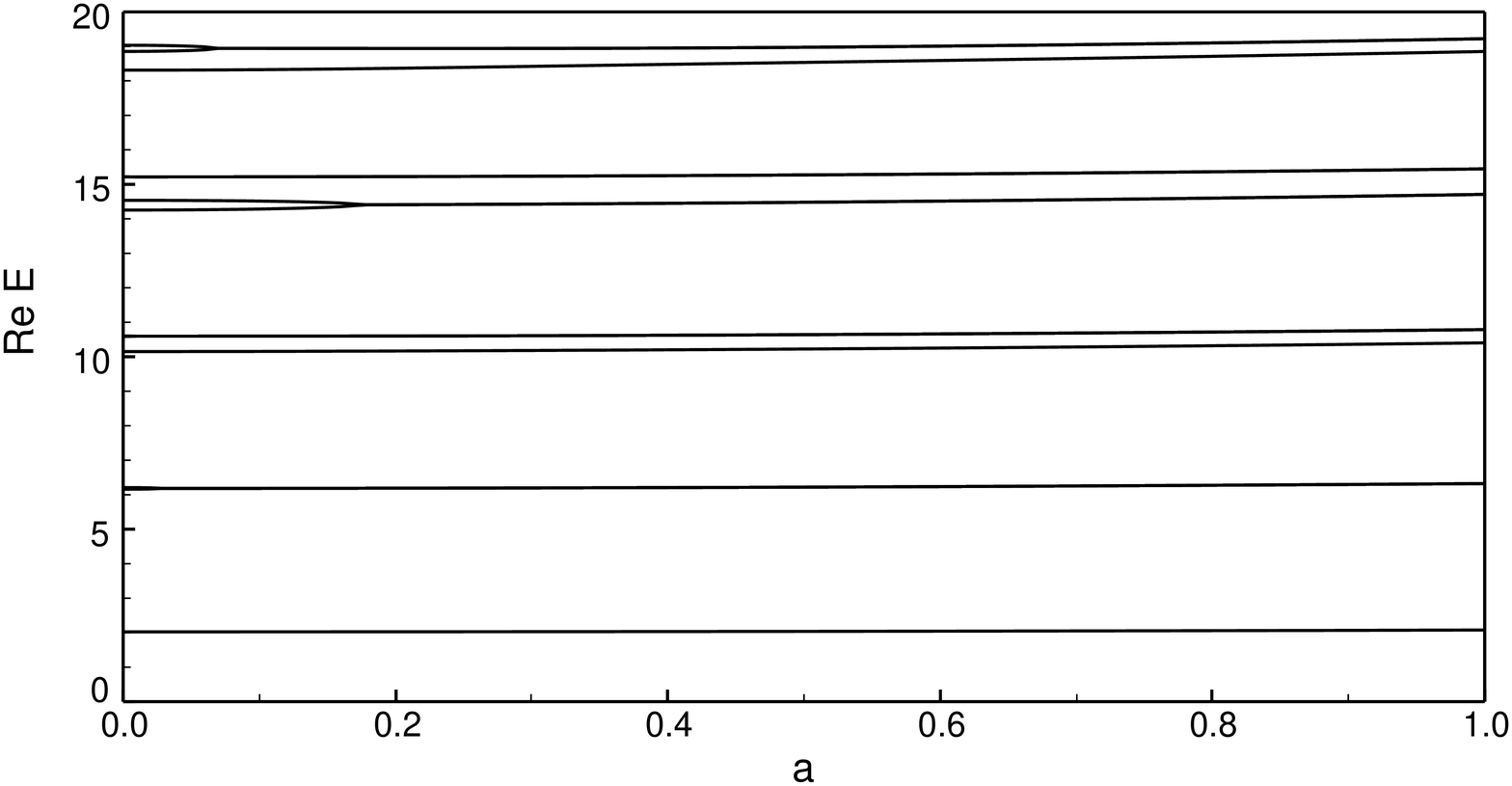} %
\includegraphics[width=6cm]{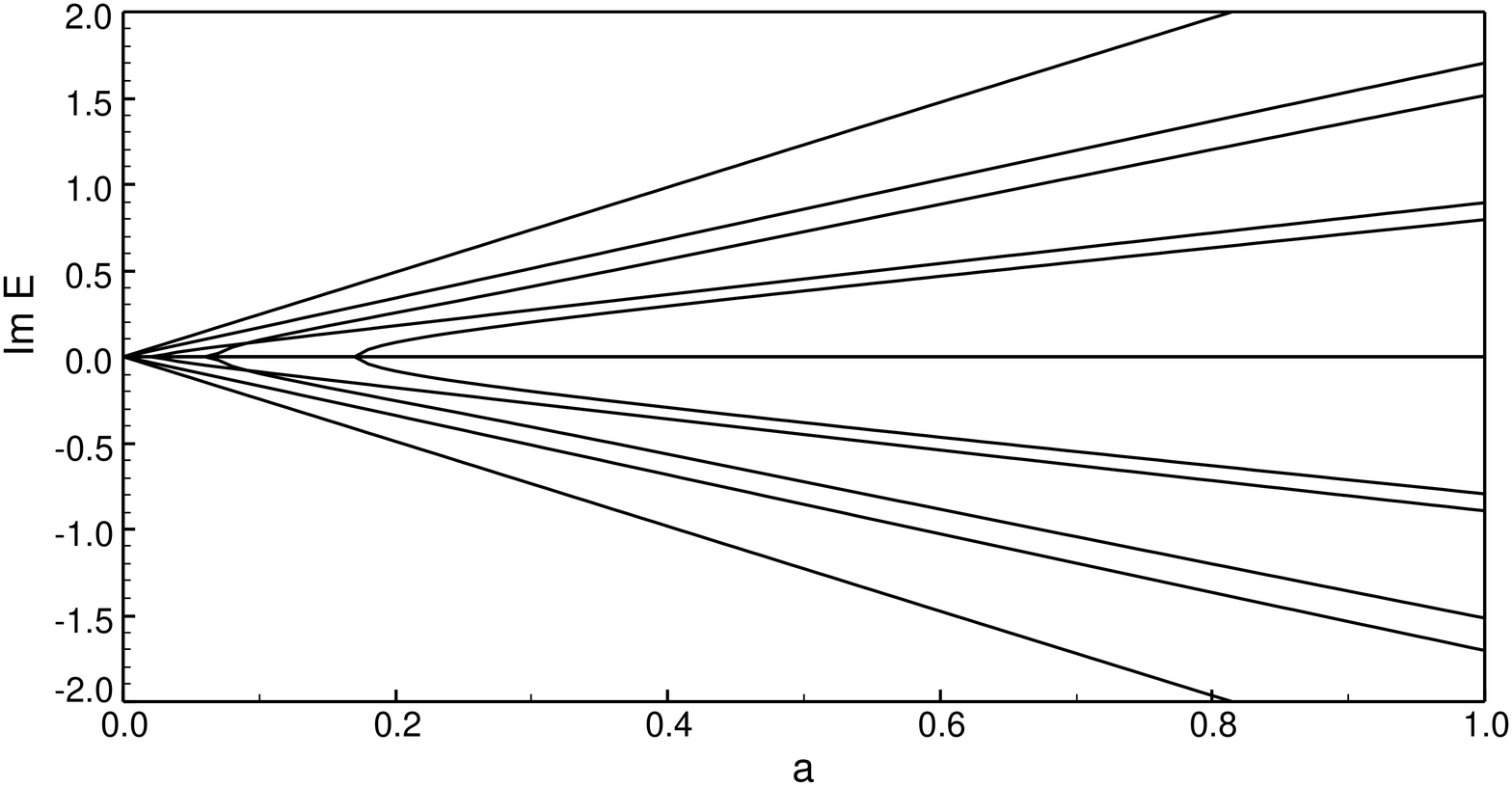} %
\includegraphics[width=6cm]{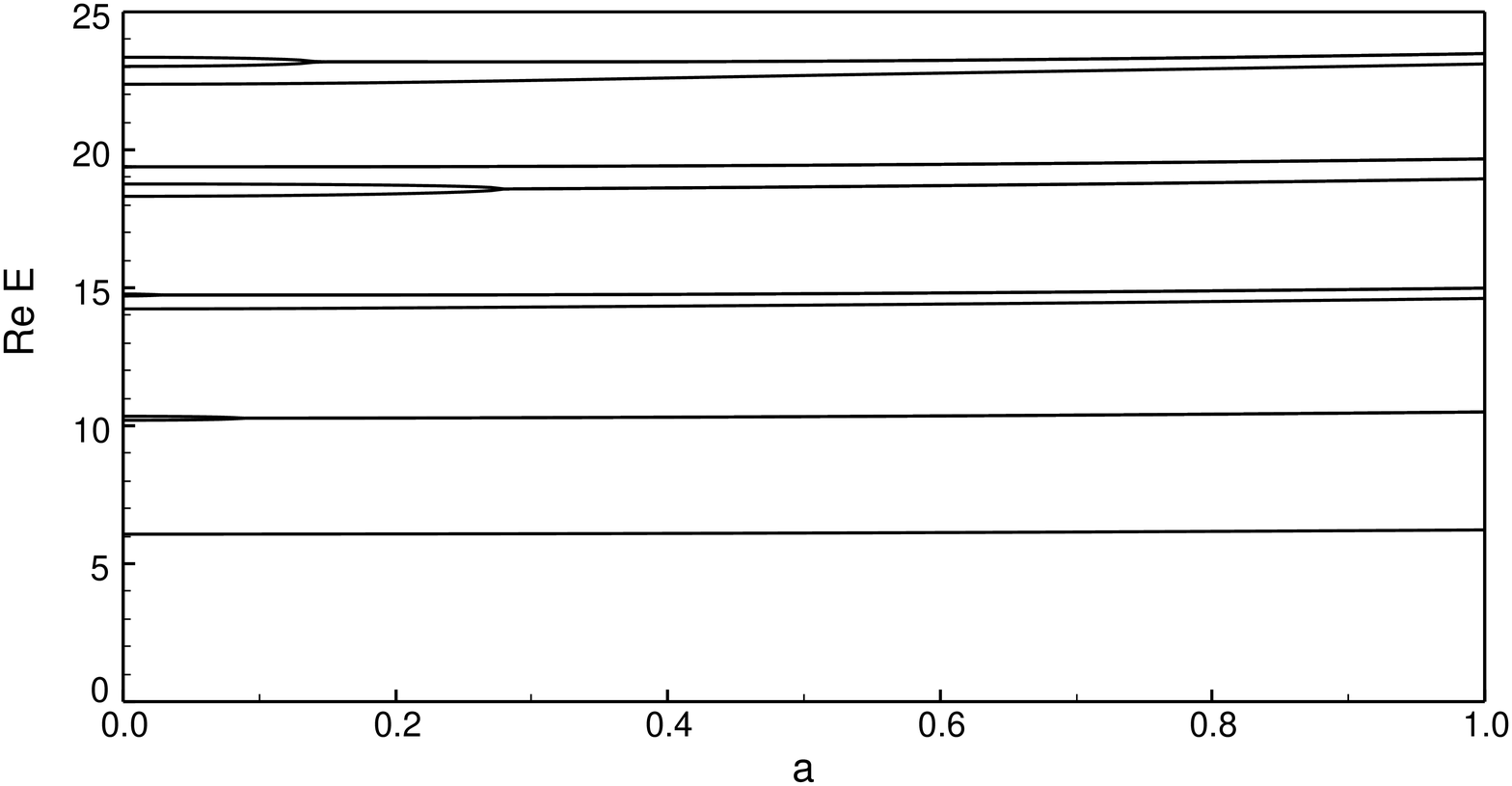} %
\includegraphics[width=6cm]{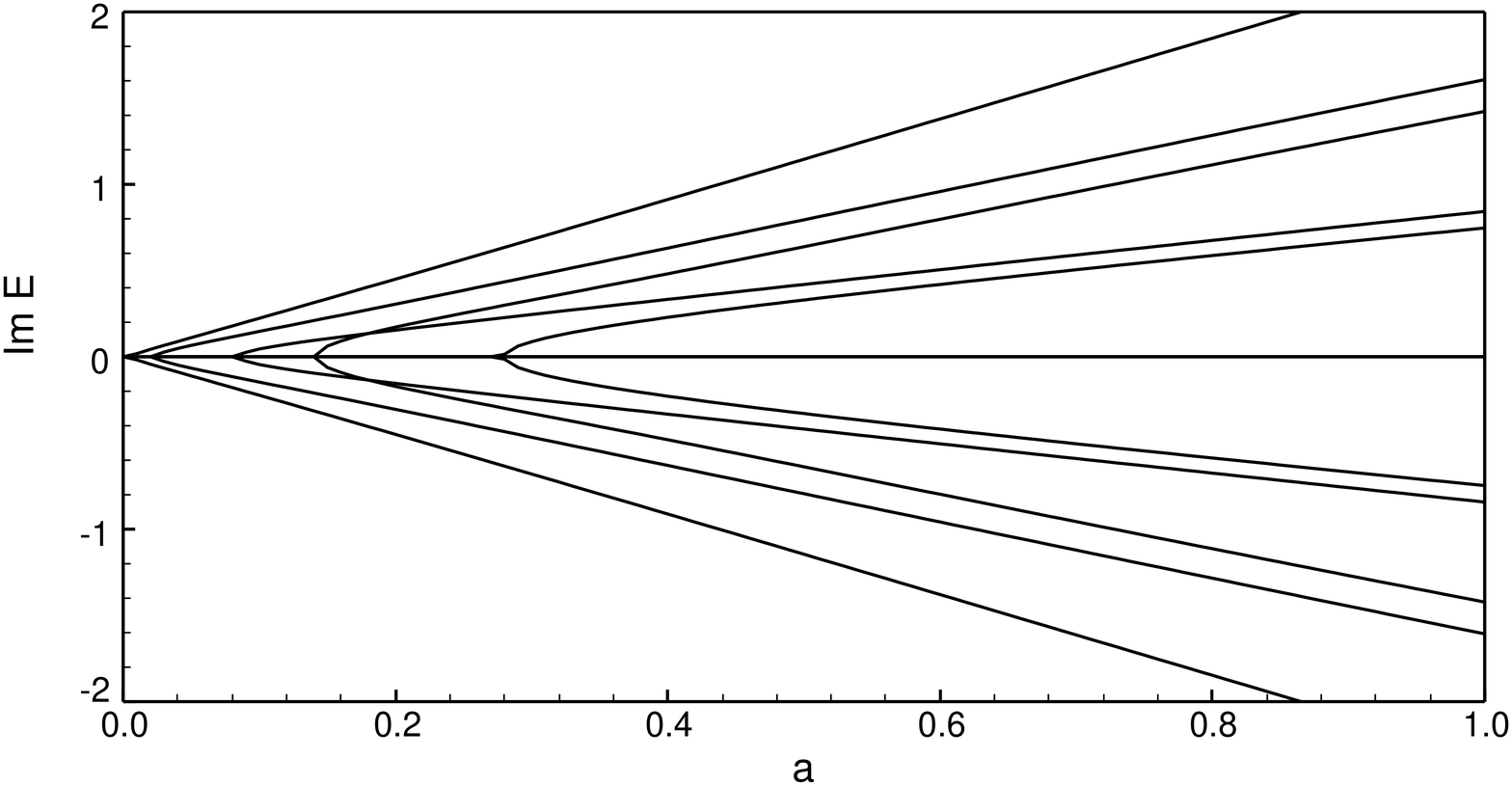} %
\includegraphics[width=6cm]{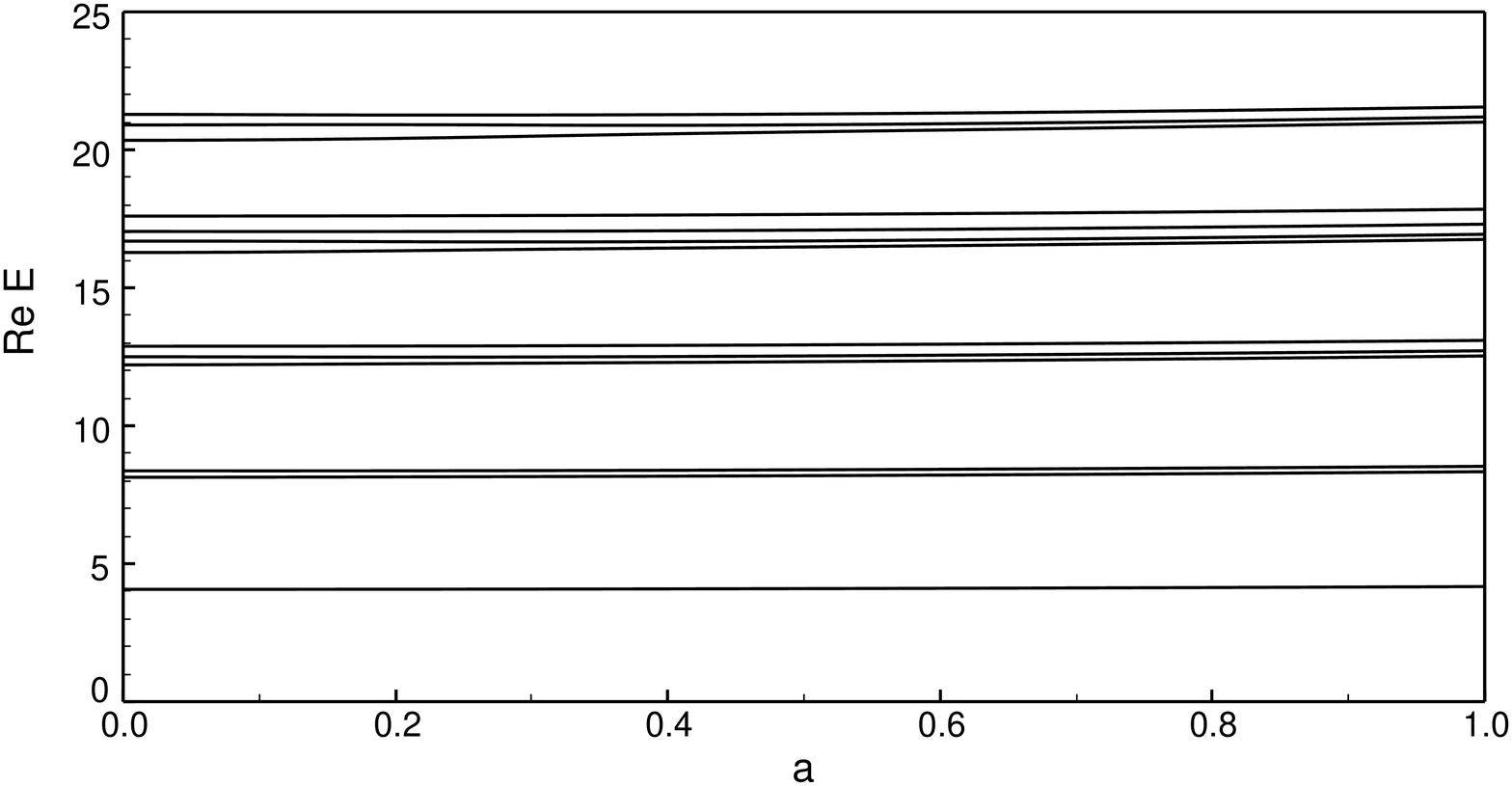} %
\includegraphics[width=6cm]{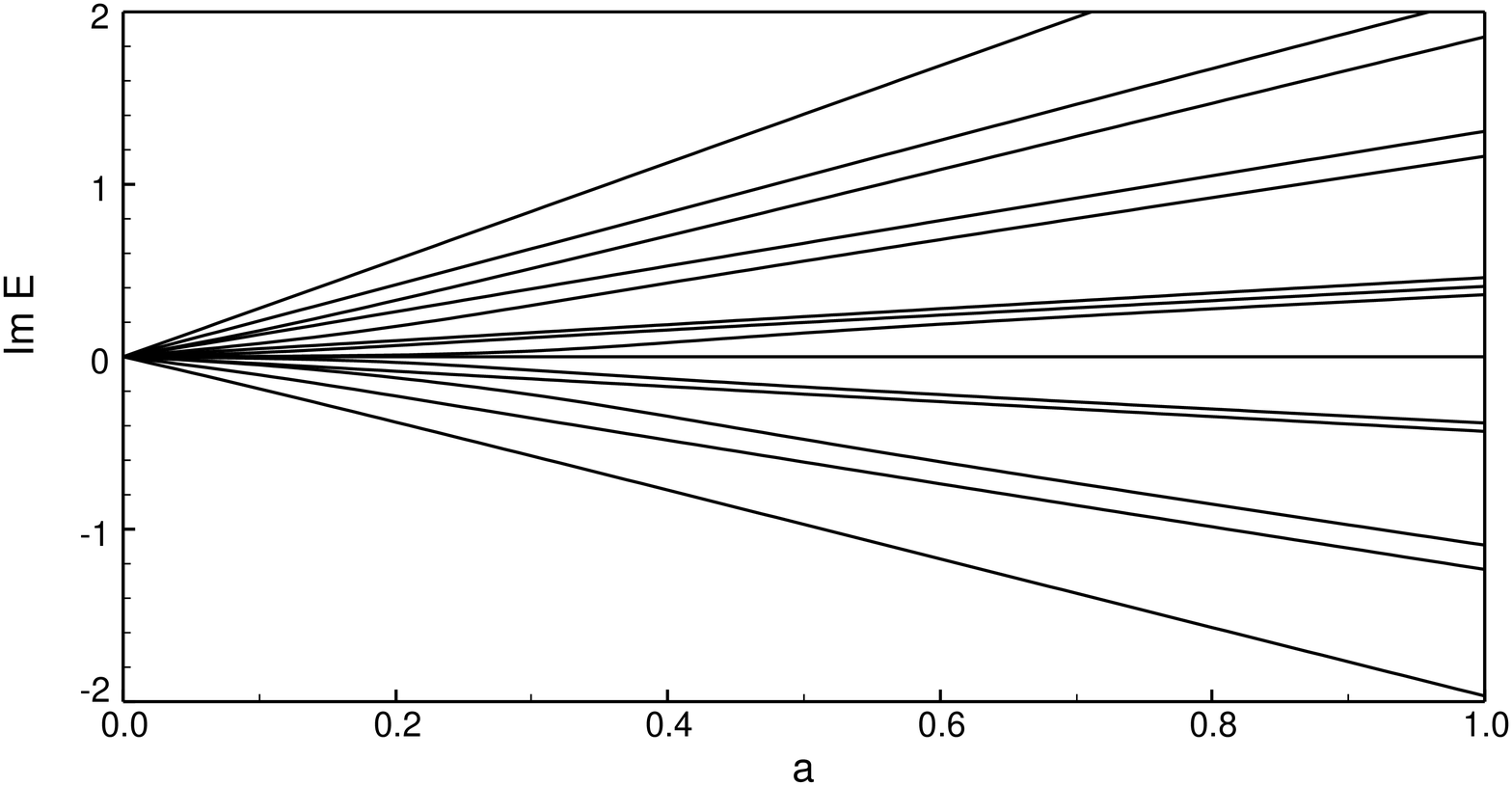} %
\includegraphics[width=6cm]{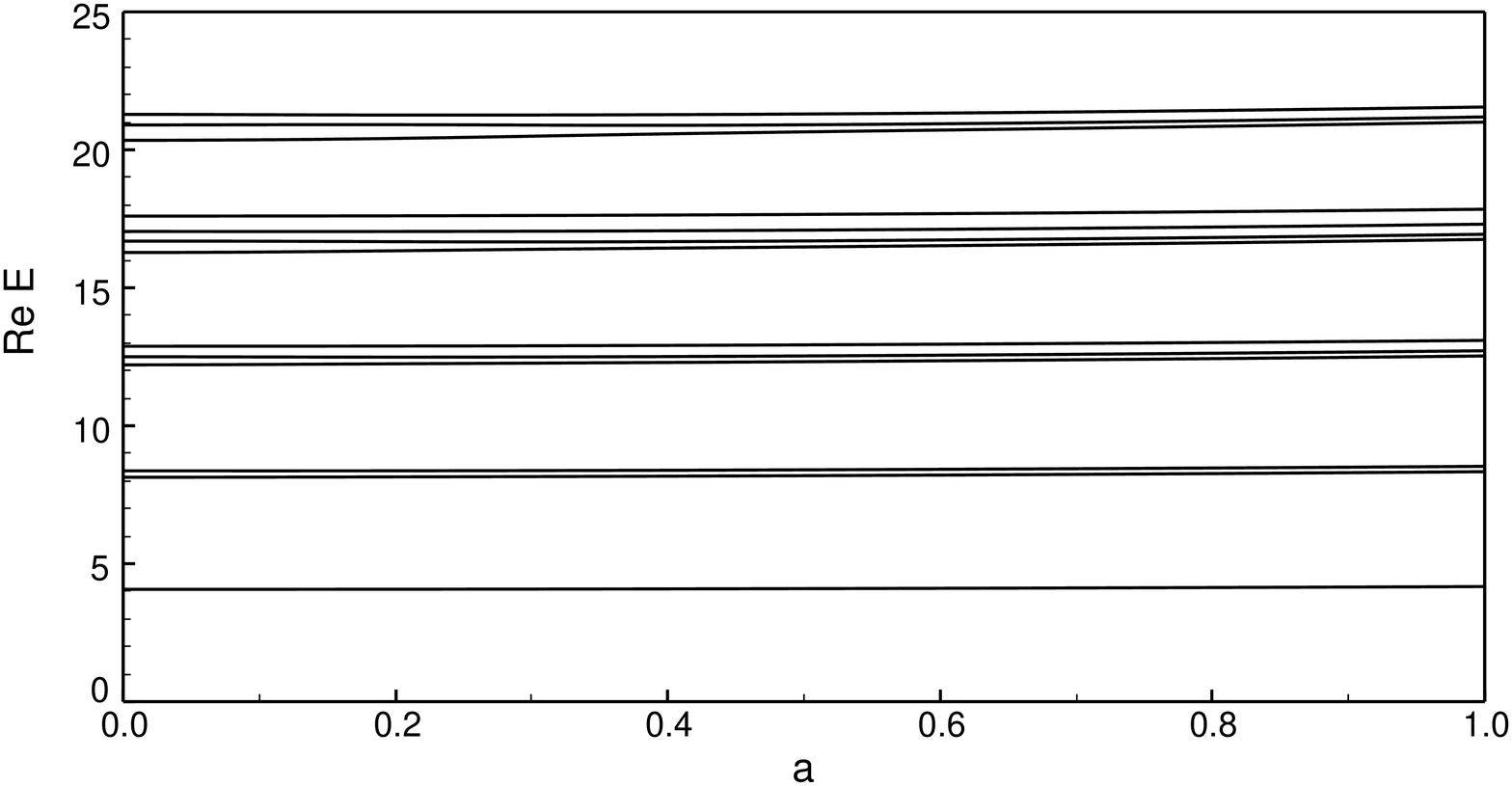} %
\includegraphics[width=6cm]{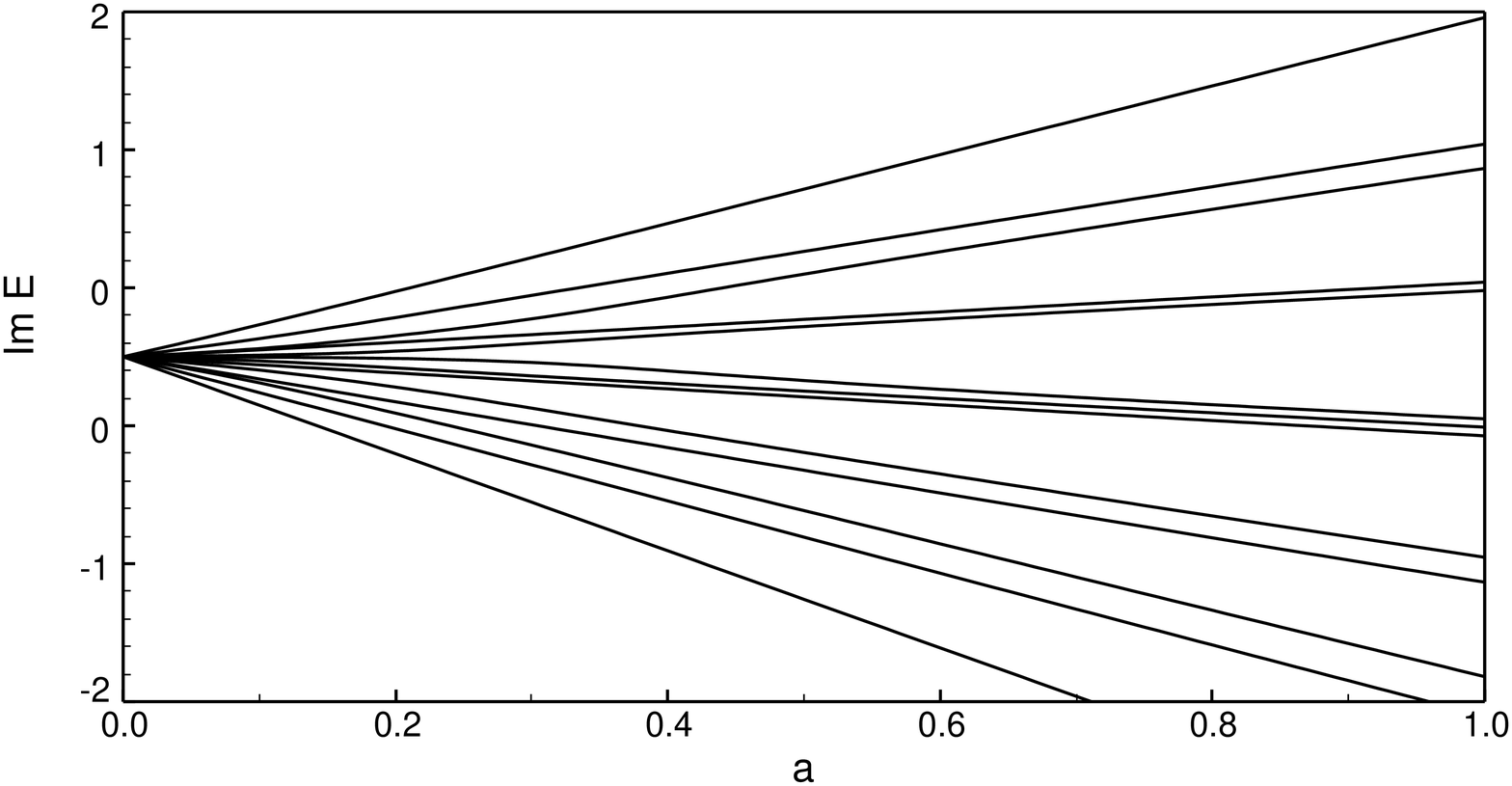}
\end{center}
\caption{Lowest eigenvalues with symmetry $A_1$, $A_2$, $B_1$ and $B_2$ (top
to bottom) of the Hamiltonian operator (\ref{eq:H_present}) with $\alpha=0$
and $\beta=0.1$}
\label{fig:Epres2}
\end{figure}

Doubts have been arisen about the existence of a discrete spectrum for the
Hamiltonian (\ref{eq:H_present}) with $\alpha =0$ and $a=0$\cite{BDMS01}.
Our numerical results suggest that it already exhibits positive discrete
spectrum. However, we have decided to choose the less controversial model
with $\alpha =1$ which enables us to obtain more accurate eigenvalues with
smaller matrix dimension. Figure~\ref{fig:Epres2} shows results for $\alpha
=1$, $\beta =0.1$ and $0\leq a\leq 1$. We appreciate that the $A$ states
exhibit phase transitions at nonzero values of $a$ but the eigenvalues of
symmetry $B$ are complex for all $a>0$ as argued above. We clearly see that
in this case the PT-symmetry is broken at $a=0$ and the phase transition
takes place at the trivial Hermitian limit. In other words, the PT phase
transition is not such a robust phenomenon as it is believed\cite{BW12}.

\section{Conclusions}

\label{sec:conclusions}

Throughout this paper we have discussed Hamiltonians that are Hermitian when
a potential parameter $a$ is zero and non-Hermitian but PT symmetric when $%
a\neq 0$. Those in sections \ref{sec:Barbanis}, \ref{sec:H-H} and \ref
{sec:H3D} discussed earlier by several authors exhibit different kinds of
PGS but they share the property of having phase transitions at nonzero
values of $a$\cite{BW12}. On the other hand, the exactly solvable
PT-symmetric harmonic oscillator of section \ref{sec:solvable} exhibits a
phase transition at $a=0$; that is to say, some of its eigenvalues are
complex for all values of $a>0$. This operator exhibits $C_{2v}$ PGS and the
eigenvalues for the $B$ eigenfunctions are complex. For such eigenfunctions
the PT symmetry is broken for all values of $a$ and the phase transition
occurs at the Hermitian limit.

In order to verify if the broken PT symmetry was due to PGS and not to the
particular form of the Hamiltonian (an exactly solvable two-dimensional
harmonic oscillator) we constructed other simple but nontrivial examples
with the same PGS and found exactly the same behaviour: the eigenvalues with
eigenfunctions of symmetry $B$ are complex for all nonzero values of the
model parameter $a$. It is likely that broken PT symmetry may also be
associated to other PGS. The most important conclusion of this paper is that
the existence of a phase transition as a high-energy phenomenon\cite{BW12}
is not a general property of PT-symmetric multidimensional oscillators. It
does not appear to be a robust phenomenon.

\begin{thebibliography}{99}
\bibitem{CGM80}  E. Caliceti, S. Graffi, and M. Maioli, Perturbation theory
of odd anharmonic oscillators, Commun. Math. Phys. 75 (1980) 51-66.

\bibitem{A95}  G. Alvarez, Bender-Wu branch points in the cubic oscillator,
J. Phys. A 28 (1995) 4589-4598.

\bibitem{BB98}  C. M. Bender and S. Boettcher, Real Spectra in Non-Hermitian
Hamiltonians Having PT Symmetry, Phys. Rev. Lett. 80 (1998) 5243-5246.

\bibitem{B07}  C. M. Bender, Making sense of non-Hermitian Hamiltonians,
Rep. Prog. Phys. 70 (2007) 947-1018.

\bibitem{FGZ98}  F. M. Fern\'{a}ndez, R. Guardiola, and M. Znojil,
Strong-coupling expansions for the PT-symmetric oscillators $%
V(x)=a(ix)+b(ix)^2+c(ix)^3$, J. Phys. A 31 (1998) 10105-10112.

\bibitem{FGRZ99}  F. M. Fern\'{a}ndez, R. Guardiola, J. Ros, and M. Znojil,
A family of complex potentials with real spectrum, J. Phys. A 32 (1999)
3105-3116.

\bibitem{BDMS01}  C. M. Bender, G. V. Dunne, P. N. Meisinger, and M. Simsek,
Quantum complex H\'{e}non-Heiles potentials, Phys. Lett. A 281 (2001)
311-316.

\bibitem{NA02}  A. Nanayakkara and C. Abayaratne, Semiclassical quantization
of complex Henon-Heiles systems, Phys. Lett. A 303 (2002) 243-248.

\bibitem{N02}  A. Nanayakkara, Real eigenspectra in non-Hermitian
multidimensional Hamiltonians, Phys. Lett. A 304 (2002) 67-72.

\bibitem{N05}  A. Nanayakkara, Comparison of quantal and classical behavior
of PT-symmetric systems at avoided crossings, Phys. Lett. A 334 (2005)
144-153.

\bibitem{BTZ06}  H. B\'{i}la, M. Tater, and M. Znojil, Comment on:
"Comparison of quantal and classical behavior of PT-symmetric systems at
avoided crossings" [Phys. Lett. A 334 (2005) 144], Phys. Lett. A 351 (2006)
452-456.

\bibitem{W09}  Q-H Wang, Level crossings in complex two-dimensional
potentials, Pramana J. Phys. 73 (2009) 315-322.

\bibitem{CIN10}  F. Cannata, M. V. Ioffe, and D. N. Nishnianidze, Exactly
solvable nonseparable and nondiagonalizable two-dimensional model with
quadratic complex interaction, J. Math. Phys. 51 (2010) 022108.

\bibitem{BW12}  C. M. Bender and D. J. Weir, PT phase transition in
multidimensional quantum systems, J. Phys. A 45 (2012) 425303.

\bibitem{HV13}  C. R. Handy and D. Vrincenau, Orthogonal polynomial
projection quantization: a new Hill determinant method, J. Phys. A 46 (2013)
135202.

\bibitem{PE81a}  R. A. Pullen and A. R. Edmonds, Comparison of classical and
quantal spectra for a totally bound potential, J. Phys. A 14 (1981)
L477-L484.

\bibitem{PE81b}  R. A. Pullen and A. R. Edmonds, Comparison of classical and
quantal spectra for the H\'{e}non-Heiles potential, J. Phys. A 14 (1981)
L319-L327.

\bibitem{HS90}  W. D. Heiss and A. L. Sannino, Avoided level crossing and
exceptional points, J. Phys. A 23 (1990) 1167-1178.

\bibitem{H00}  W. D. Heiss, Repulsion of resonance states and exceptional
points, Phys. Rev. E 61 (2000) 929-932.

\bibitem{HH01}  W. D. Heiss and H. L. Harney, The chirality of exceptional
points, Eur. Phys. J. D 17 (2001) 149-151.

\bibitem{H04}  W. D. Heiss, Exceptional points - their universal occurrence
and their physical significance, Czech. J. Phys. 54 (2004) 1091-1099.

\bibitem{FG13}  F. M. Fern\'{a}ndez and J. Garcia, Critical parameters for
non-hermitian Hamiltonians, arXiv:1305.5164 [math-ph]

\bibitem{W60}  E. Wigner, Normal Form of Antiunitary Operators, J. Math.
Phys. 1 (1960) 409-413.

\bibitem{T64}  M. Tinkham, Group Theory and Quantum Mechanics, (McGraw-Hill
Book Company, New York, 1964).

\bibitem{C90}  F. A. Cotton, Chemical Applications of Group Theory, (John
Wiley \& Sons, New York, 1990).

\bibitem{P65}  C. E. Porter, Fluctuations of quantal spectra, in: C. E.
Porter (Ed.), Statistical theories of spectra: fluctuations, Vol. Academic
Press Inc., New York and London, 1965.

\bibitem{F13}  F. M. Fern\'{a}ndez, On the real matrix representation of
PT-symmetric operators, arXiv:1301.7639v3 [quant-ph]
\end{thebibliography}
\end{document}